\documentclass[prb,aps,eqsecnum,twocolumn,floats,nofootinbib]{revtex4}

\usepackage{amsmath,amssymb,dsfont,graphicx,psfrag}

\newcommand{\nn}{\nonumber}
\newcommand{\hp}{(C$_5$H$_{12}$N)$_2$CuBr$_4$}

\begin{document}
\title{
Spin-spin correlations of the spin-ladder compound
(C$_5$H$_{12}$N)$_2$CuBr$_4$ measured by magnetostriction and
comparison to Quantum Monte Carlo results}

\author{Fabrizio Anfuso}
\author{Markus Garst}
\author{Achim Rosch}
\affiliation{Institut f\"ur
  Theoretische Physik, Universit\"at zu K\"oln, Z\"{u}lpicher Str. 77,
  50937 K\"oln, Germany}
\author{Oliver Heyer}
\author{Thomas Lorenz}
\affiliation{$ $II. Physikalisches Institut, Universit\"{a}t zu
K\"{o}ln, Z\"{u}lpicher Str. 77, 50937 K\"{o}ln, Germany}
\author{Christian R\"{u}egg}
\affiliation{London Centre for Nanotechnology and Department of
Physics and Astronomy, University College London, London WC1E 6BT,
United Kingdom}
\author{Karl Kr\"amer}
\affiliation{Department of Chemistry and Biochemistry, University
of Bern, Freiestrasse, CH--3000 Bern 9, Switzerland}

\date{\today}

\begin{abstract}
Magnetostriction and thermal expansion of the spin-ladder compound
piperidinium copper bromide (C$_5$H$_{12}$N)$_2$CuBr$_4$ are
analyzed in detail. We find perfect agreement between experiments
and the theory of a two-leg spin ladder Hamiltonian for more than
a decade in temperature and in a wide range of magnetic fields.
Relating the magnetostriction along different crystallographic
directions to two static spin-spin correlation functions, which we
compute with Quantum Monte Carlo, allows us to reconstruct the
magnetoelastic couplings of (C$_5$H$_{12}$N)$_2$CuBr$_4$. We
especially focus on the quantum critical behavior near the two
critical magnetic fields $H_{c1}$ and $H_{c2}$, which is
characterized by strong singularities rooted in the low
dimensionality of the critical spin-system. Extending our
discussion in Lorenz {\it et al.} [Phys.\ Rev.\ Lett., {\bf 100},
067208 (2008)], we show explicitly that the thermal expansion near
the upper critical field $H_{c2}$ is quantitatively described by a
parameter-free theory of one-dimensional, non-relativistic
Fermions. We also point out that there exists a singular quantum
critical correction to the elastic moduli. This correction is
proportional to the magnetic susceptibility $\chi$ which diverges
as $\chi \sim 1/\sqrt{T}$ at the critical fields and thus leads to
a strong softening of the crystal.
\end{abstract}


\maketitle

\section{Introduction}

By now, the field-induced Bose-Einstein condensation of
magnons\cite{Batyev84,Giamarchi99} and its one-dimensional
analog\cite{Affleck90,Affleck91} have been observed in many
different spin compounds\cite{Review} like coupled spin-dimer
systems,\cite{Ruegg03,Johannsen05,Lorenz07} arrays of coupled
spin-1 chains\cite{Zapf1, Zapf2} or spin
ladders.\cite{watson01a,Garlea} Typically, the ground state of
these systems is a spin singlet, separated by a finite energy gap
from the lowest triplet excitation. Increasing the magnetic field,
two quantum phase transitions are induced. There is a first
quantum critical point at a field $H_{c1}$ where the spin gap
closes, and a second at $H_{c2}>H_{c1}$, when the fully
field-polarized state becomes the exact  ground state of the
system. The phase between the two critical fields is gapless and,
dependent on the dimensionality, is characterized either by
long-range order or power-law
correlations\cite{Schulz86,Affleck91}  among magnetic moments.

Recently, we have reported in Ref.~\onlinecite{ThomasPRL}
high-resolution measurements of thermal expansion and
magnetostriction of single crystalline piperidinium copper bromide
\hp. This compound has been previously identified\cite{watson01a}
to be a good realization of a two-leg spin ladder. As such, it
serves as a perfect model system to test experimentally the
validity and applicability of theoretical predictions for quantum
critical thermodynamics in a controlled fashion. In particular, it
has been argued that the thermal expansion, $\alpha$, is an
especially useful probe in the presence of pressure coupling to
the critical subsystem because $\alpha$ is then more singular than
the specific heat. This results in a divergent Gr\"uneisen
parameter close to quantum
criticality\cite{Lorenz07,Zhu03,Kuechler03,stark07} with an
exponent characteristic for the universality class of the
transition. Furthermore, as the thermal expansion is given by the
entropy derivative with respect to pressure, $\alpha \propto
\partial S/\partial p$, it should exhibit a characteristic sign
change near the quantum phase transition signaling the
accumulation of entropy.\cite{Garst05} The experimental check of
these predictions in a controlled model system gives important
insights about their reliability and applicability in the pursuit
of quantum criticality, especially in materials where the nature
of the quantum critical point is
unknown.\cite{HeavyFermions,HeavyFermions2}

In \hp,  the presence of two adjacent quantum critical points,
$H_{c1}$ and $H_{c2}$, indeed leads to a rich structure of sign
changes of thermal expansion reflecting the positions of entropy
extrema in the phase diagram. As illustrated in Ref.~\onlinecite{ThomasPRL},
at low temperatures a maximum in entropy is located near each of
the critical fields, $H_{c1/2}$, with an enclosed minimum in between.
Upon increasing temperature $T$, the two maxima approach each other
and merge at higher temperatures. Moreover, the low dimensionality of
the spin-ladder 
results in a diverging thermal expansion\cite{ThomasPRL} behaving
as $\alpha \sim 1/\sqrt{T}$ at the two quantum critical points.

The compound \hp\ has a monoclinic crystal structure ($\beta
\simeq 99.3^\circ$)~\cite{Patyal90} with the legs of the spin
ladders oriented along the $a$ axis and the rungs roughly
($\approx 20^\circ$) along the $c^\star$ axis of the reciprocal
lattice. The single crystals used for our study have been grown by
slow evaporation of a solution of \hp\ in ethanol. Typical
crystals grow in plates of several $mm^3$ with (010) faces and the
$a$ and $c$ axes oriented parallel to the edges.\cite{ruegg} The
measurements have been performed on a home-built capacitance
dilatometer in longitudinal magnetic fields up to 17~T for
temperatures $0.3~{\rm K} \lesssim T \lesssim 10$~K. The absence
of three-dimensional Ne\'el order in \hp\ down to temperatures
$T_N < 100$~mK implies a very weak inter-ladder coupling giving
rise to an extended temperature regime controlled by
one-dimensional physics.\cite{ruegg} In this regime, the magnetic
subsystem of \hp\ is well-described by the two-leg spin ladder
Hamiltonian
\begin{align} \label{modelhamiltonian}
\mathcal{H} &= \sum_{i=1}^N \left[
J_\perp \mathbf{S}_{i,1} \mathbf{S}_{i,2}
+ J_\parallel
\left(\mathbf{S}_{i,1} \mathbf{S}_{i+1,1} + \mathbf{S}_{i,2}
\mathbf{S}_{i+1,2} \right)
\right.
\nn\\ &\hspace{3em}
\left.
- g \mu_B H \left(S^z_{i,1} + S^z_{i,2} \right)\right],
\end{align}
where the first and second index specify the rung and the chain,
respectively. We have identified the exchange couplings to be
$J_\perp/k_B = 12.9$~K and $J_\parallel/k_B = 3.6$~K, see
Sec.~\ref{determinationJ}. The $g$-factor is in fact a tensor and
so depends on the crystallographic axis along which the magnetic
field $H$ is applied.~\cite{Patyal90}

In the following, we shortly review the properties of the
different ground states of the spin-ladder Hamiltonian
(\ref{modelhamiltonian}) that appear as a function of magnetic
field.\cite{Chitra96} At zero magnetic field, $H=0$, the ground
state of the spin ladder is a singlet made of short-ranged valence
bonds.\cite{Haldane83} Due to the dominant rung interaction
$J_\perp/J_\parallel \sim 4$, these valence bonds can be pictured
in a good approximation as the singlet states of the spin dimers
located on each rung of the
ladder.\cite{Chaboussant98,Mila98,Totsuka98} The resulting
spectrum has a finite gap separating the ground state from the
lowest-lying triplet excitations, and, as a consequence, the
spin-spin correlations decay exponentially with distance. Upon
increasing the magnetic field, this gap decreases until the
lowest-lying dispersing triplet excitation touches the singlet
energy giving rise to a quantum phase transition at $H_{c1}$. For
higher fields, singlet and triplet excitations hybridize giving
rise to a finite magnetization in the ground state. This phase is
a Luttinger liquid and has gapless excitations representing
fluctuations of the magnetization perpendicular to the applied
magnetic field. Finally, for even higher magnetic fields the zero
temperature magnetization saturates above a critical field, i.e.\
for $H>H_{c2}$. Here, the excitations on top of the fully
polarized ground state have again a gap that increases with the
distance to the transition $H-H_{c2}$.

Whereas in Ref.~\onlinecite{ThomasPRL} we focused on the thermal
expansion and magnetostriction along the $c^\star$ axis of \hp, in
the present article we also discuss our experimental data measured
along the $a$ and $b$ axes. Furthermore, we present a detailed
comparison with results of Quantum Monte Carlo (QMC) simulations
of the Hamiltonian (\ref{modelhamiltonian}) for the full magnetic
field range extending beyond the second critical field $H_{c2}$.
As explained in detail in Sec.~\ref{sec:Dilatometry}, we
numerically evaluate for this analysis spin-spin correlation
functions that are related to magnetostriction, as was noted
before by Zapf {\it et al.} in Ref.~\onlinecite{Zapf07}. We find
excellent quantitative agreement between theory and experiment
that allows to determine the magnetoelastic couplings of \hp\ and
their respective uniaxial pressure dependencies. In addition, we
present a discussion of the quantum critical thermal expansion
near $H_{c2}$ in Sec.~\ref{sec:CriticalThermalExp}. The effective
critical theory describing its behavior is known exactly, which
enables us to directly calculate the thermal expansion near
$H_{c2}$ without adjustable parameters. Finally, in
Sec.~\ref{sec:FirstOrder} we note that not only the thermal
expansion but also the elastic moduli obtain a quantum critical
correction that diverges upon approaching the critical fields.
This results in a strong softening of the crystal that finally
triggers a first-order transition in the elastic system. We
present a mean-field discussion of the phase diagram with the
expected position of the line of first-order transitions and the
accompanied coexistence regions. Finally, we estimate that this
effect is too weak in \hp\ to be observable because the strong
one-dimensional signatures will be cut-off by inter-ladder
interactions before a first-order transition can develop.

\section{Magnetoelastic coupling in $\rm\bf (C_5H_{12}N)_2CuBr_4$}
\label{sec:Dilatometry}

In the range of temperatures considered here, the main
contribution to thermodynamics  of \hp\ can be attributed to the
spin system, i.e., the spin ladders. As the inter-ladder coupling
is sufficiently weak, $T_N < 100$~mK, it can be neglected, and we
can approximate the magnetic part of the free energy,
\begin{align} \label{FreeEnergy}
F_m = - k_B T \ln {\rm  tr} \{e^{-\hat{H}/k_B T}\},
\end{align}
as arising from an ensemble of decoupled spin-ladders. The
Hamiltonian thus decomposes into a sum of Hamiltonians, $\hat{H} =
\sum_{n} \mathcal{H}_{n}$, where the index $n$ counts the number
of equivalent ladders in the transverse plane orthogonal to the
ladder-leg direction, the $a$ axis in \hp. The Hamilonian
$\mathcal{H}_n$ represents a single ladder system as defined in
(\ref{modelhamiltonian}). The free energy density is
\begin{align} \label{FreeEnergySingleLadder}
\frac{1}{V} F_m =  \frac{1}{V_D N} \mathcal{F}_m,
\end{align}
where $V_D = 859$~\AA$^3$ is the volume per rung,~\cite{ruegg} and
$N$ is the number of rungs in the ladder. The free energy
$\mathcal{F}_m$ deriving from a single spin-ladder Hamiltonian
$\mathcal{H}$, (\ref{modelhamiltonian}), is given by
$\mathcal{F}_m = - k_B T \ln {\rm  tr}\{e^{-\mathcal{H}/k_B T}\}$.

The dilatometric properties of \hp\ are characterized by its
elastic energy,\cite{LandauElasticity} whose low-temperature and
magnetic-field dependencies we expect to be dominated by the spin
subsystem. The couplings, $J_\alpha$, with $\alpha =
\perp,\parallel$, of the spin-ladder Hamiltonian
(\ref{modelhamiltonian}), are determined by the exchange integrals
computed from the electronic wavefunctions for given values of
lattice parameters. In the presence of strain in the lattice,
$u_{ij}$, the lattice parameters slightly change and thus induce a
variation in the effective couplings, $J_{\alpha} =
J_{\alpha}(u_{ij})$. Expanding this dependence to linear order in
the strain, $J_{\alpha}(u_{ij}) \approx J_{\alpha} +
g_{\alpha}^{ij} u_{ij}$, with $g_{\alpha}^{ij} = g_{\alpha}^{ji} =
\partial J_{\alpha}/ \partial u_{ij}|_{u=0}$, we can separate from
(\ref{modelhamiltonian}) a magnetoelastic interaction Hamiltonian,
\begin{align} \label{SpinStrain}
\lefteqn{\mathcal{H}_{\rm int} =}
\\\nn
&\sum_{i=1}^N \left[
g^{nm}_\perp \mathbf{S}_{i,1} \mathbf{S}_{i,2}
+ g^{nm}_\parallel
\left(\mathbf{S}_{i,1} \mathbf{S}_{i+1,1} + \mathbf{S}_{i,2}
\mathbf{S}_{i+1,2} \right)
\right]  u_{nm},
\end{align}
where summation over the indices $n$ and $m$ is implied. The
strain $u_{nm}$ fluctuates locally and so depends on the site
index $i$. The spin-ladder couples linearly to strain and thus
acts as a force on the lattice. The lattice responds according to
Hooke's law and, as a consequence, inherits the characteristic
temperature and magnetic field dependence of the spin-spin
correlations in the ladder.

\begin{figure}
\includegraphics[width= 0.3\linewidth]{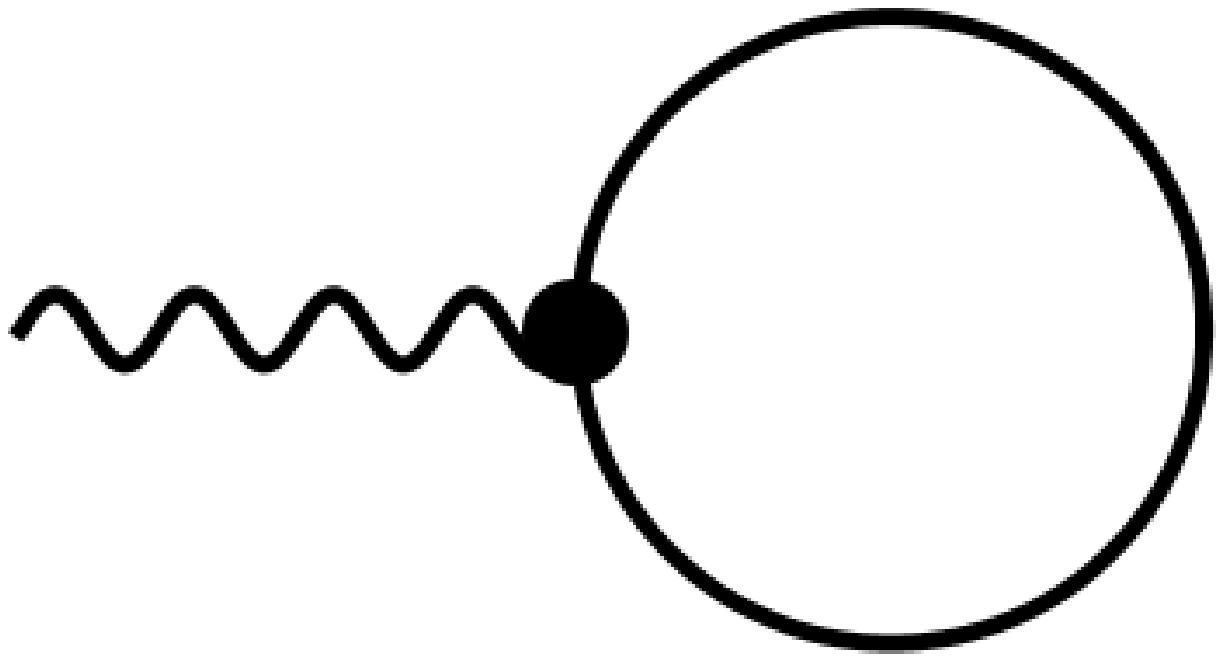}
\hspace{0.15\linewidth}
\includegraphics[width= 0.5\linewidth]{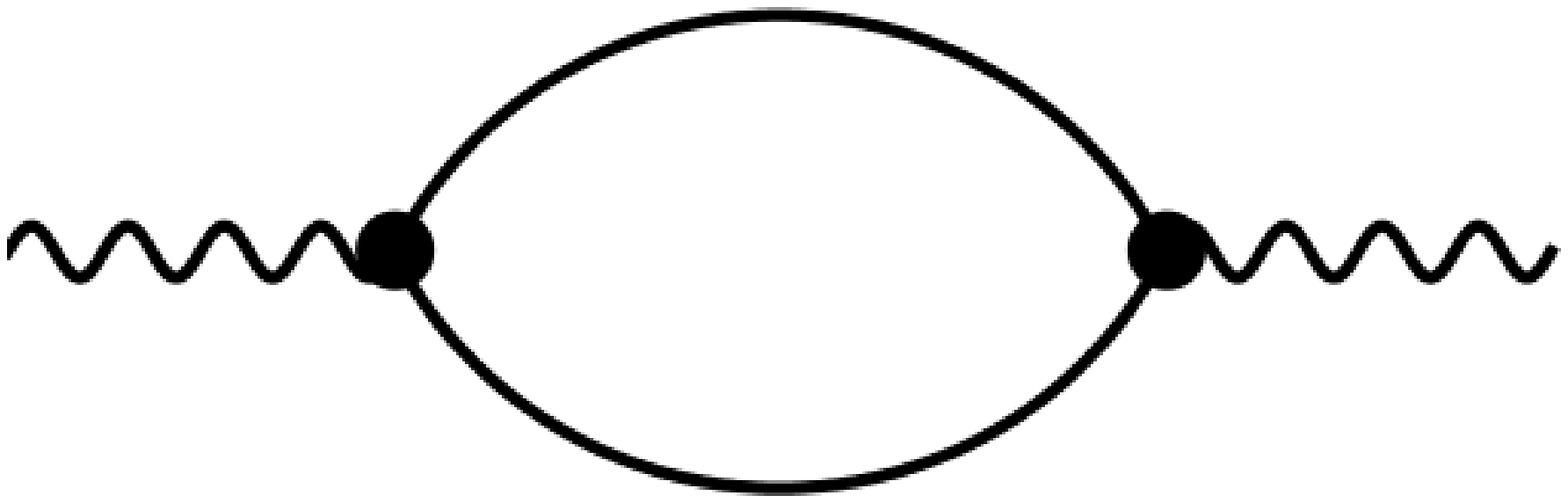}
\\
(a) \hspace{0.7\linewidth} (b)
\caption{\label{fig:MagnetoelasticCoupling} Pictorial
representation of the magnetoelastic mode-coupling corrections to
the elastic energy, (\ref{ElasticEnergy}). The wiggled line
represents strain $u_{ij}$ and the dot is the strain-spin
interaction $g_\alpha$. The loop of straight lines in diagram (a)
represents a two- and in diagram (b) a four-spin correlation
function. The tadpole diagram (a) results in a force on the
lattice responsible for the magnetostriction and thermal expansion
effect. Diagram (b) modifies the elastic tensor. }
\end{figure}

In order to obtain an effective elastic theory, we consider the
corrections to the elastic energy density, $\mathcal{E}$, in
second-order perturbation theory in the magnetoelastic coupling
$g_\alpha$. Generally, a correction of order $g^n_\alpha$ to the
elastic energy is weighted by a $2n$-spin correlation function
that can be conveniently represented as an $n^{\rm th}$ derivative
of the free energy $F_m$, Eq.~(\ref{FreeEnergy}), with respect to
the couplings $J_\alpha$; the most important contributions are
illustrated in Fig.~\ref{fig:MagnetoelasticCoupling}. The
resulting effective elastic energy density reads
\begin{align} \label{ElasticEnergy}
\mathcal{E}(u) = \frac{1}{2} u_{ij} c_{ijkl} u_{kl}
+ \frac{1}{V}\frac{\partial F_m}{\partial J_\alpha}
g_{\alpha}^{ij}  u_{ij} + \mathcal{O}(g^3).
\end{align}
The magnetoelastic coupling gives a contribution linear in strain
$u_{ij}$ represented by the tadpole diagram (a) of
Fig.~\ref{fig:MagnetoelasticCoupling}. The derivative $\partial
F_m/\partial J_\alpha$ acts on the lattice as a $H$ and $T$
dependent force that will lead to magnetostriction and thermal
expansion. In addition, the elastic tensor obtains a correction
that is second order in the interaction $g$, see
Fig.~\ref{fig:MagnetoelasticCoupling}b
\begin{align} \label{ElasticTensor}
c_{ijkl} = c^{0}_{ijkl} +  \frac{1}{V}\frac{\partial^2 F_m}
{\partial J_\alpha \partial J_\beta}g_{\alpha}^{ij} g_\beta^{kl}.
\end{align}
The tensor $c^0_{ijkl}$ characterizes the elasticity of the
crystal in the absence of spin-strain coupling. Typically, the
magnetoelastic coupling $g_\alpha$ is small and the second-order
correction in (\ref{ElasticTensor}) can be neglected. In the
following, we will therefore treat the elastic tensor $c_{ijkl}$
as independent of magnetic field and temperature. As we will
discuss in Sec.~\ref{sec:FirstOrder}, this assumption is very well
justified in the temperature range of the experimental data that
we present below. Only very close to the quantum critical points,
$H_{c1/c2}$, the tensor $c_{ijkl}$ obtains a strong dependence on
$H$ and $T$ as the second-order derivatives $\partial^2
F_m/\partial J_\alpha \partial J_\beta$ diverge, resulting in a
pronounced softening of the crystal.

Treating the effective elastic theory (\ref{ElasticEnergy}) on a
mean-field level, we can minimize the elastic energy
(\ref{ElasticEnergy}), $\partial \mathcal{E}/\partial u_{ij}=0$,
to obtain the strain
\begin{align} \label{MeanStrain}
u_{ij}
 = - (c^{-1})_{ijkl} \frac{1}{V}\frac{\partial F_m}{\partial J_\alpha}
 g_{\alpha}^{kl}
  = \mathcal{S}_\alpha \gamma^{ij}_\alpha,
\end{align}
where we made use of the inverse of the elastic tensor,
$(c^{-1})_{ijkl} c_{klnm} = \frac{1}{2} \left(\delta_{in}
\delta_{jm} + \delta_{im} \delta_{jn}\right)$. For later
convenience, we also introduced the dimensionless constants
\begin{align}
\gamma^{ij}_\alpha  = - (c^{-1})_{ijkl} \frac{1}{V_D}
g_\alpha^{kl} = - \left.(c^{-1})_{ijkl} \frac{1}{V_D}
\frac{\partial J_{\alpha}}{\partial u_{kl}} \right|_{u=0}
\end{align}
that quantify the strain dependence of the exchange energies;
$V_D$ is again the volume per rung. Moreover, we defined
\begin{align}
\mathcal{S}_\alpha = \frac{V_D}{V}\frac{\partial F_m}{\partial J_\alpha}
= \frac{1}{N} \frac{\partial \mathcal{F}_m}{\partial J_{\alpha}},
\end{align}
see Eq.~(\ref{FreeEnergySingleLadder}). These quantities can be
identified with the spin-spin correlation functions along the rung
and along the leg of the ladder, $\alpha = \perp, \parallel$,
\begin{subequations}
\label{Sfunctions}
\begin{align}
\mathcal{S}_\perp(T,h) &=  \frac{1}{N}
\sum^N_{i} \langle \mathbf{S}_{i,1} \mathbf{S}_{i,2} \rangle_\mathcal{H}
\\
\mathcal{S}_\parallel(T,h) &=  \frac{1}{N} \sum^N_{i} \langle
\mathbf{S}_{i,1} \mathbf{S}_{i+1,1} + \mathbf{S}_{i,2}
\mathbf{S}_{i+1,2} \rangle_\mathcal{H} \, ,
\end{align}
\end{subequations}
where $\langle \hat{O} \rangle_{\mathcal{H}} =$ tr$ \{ \hat{O}\,
e^{-\mathcal{H}/k_B T} \}/$tr$ \{ e^{-\mathcal{H}/k_B T}\}$. They
depend on temperature, $T$, and magnetic field, $H$; the
dependence on the latter only enters via the effective field $h =
g \mu_B H$. These correlators will be basic ingredients in what
follows. They determine the $H$ and $T$ dependence of the strain
due to the magnetoelastic coupling and lead to the
magnetostriction and the magnetic contribution to the thermal
expansion of \hp.

{\it Uniaxial length changes:} The resulting length change along
some axis $n$ can be obtained by projection of the strain
(\ref{MeanStrain}),
\begin{align} \label{UniaxialLC}
\frac{\delta L_{n}}{L_{n}} = \hat{n}_i u_{ij} \hat{n}_j =
\mathcal{S}_\perp \gamma^{n}_\perp +
\mathcal{S}_\parallel \gamma^{n}_\parallel,
\end{align}
where $\hat{n}$ is a unit vector along the $n$ axis. We introduced
the abbreviation
\begin{align} \label{MixingCoefficients0}
\gamma^{n}_\alpha = \hat{n}_i \gamma^{ij}_\alpha \hat{n}_j
=- \left.\hat{n}_i \hat{n}_j (c^{-1})_{ijkl} \frac{1}{V_D}
\frac{\partial J_{\alpha}}{\partial u_{kl}} \right|_{u=0}
\end{align}
Using $(c^{-1})_{ijkl} = \partial u_{kl}/\partial \sigma_{ij}$,
with the stress tensor $\sigma_{ij}$, we identify the coefficients
$\gamma^n_\alpha$ with the derivatives of the exchange couplings
with respect to uniaxial pressure $p_n$,
\begin{align} \label{MixingCoefficients}
\gamma^{n}_\alpha =
-\hat{n}_i  \frac{1}{V_D} \frac{\partial J_{\alpha}}{\partial \sigma_{ij}} \hat{n}_j
=
\frac{1}{V_D} \frac{\partial J_{\alpha}}{\partial p_n}.
\end{align}
Length  measurements on \hp\ were performed along the $a$, $b$ and
$c^\star$ axes. Our detailed comparison between theory and
experiment allows us to determine the coefficients
$\gamma^n_{\perp,\parallel}$ for $n=a, b, c^\star$, which will be
given below.

{\it Magnetostriction: } The magnetostriction is defined as the
change of length as a function of magnetic field $H$
\begin{align}
\frac{\Delta L_n(T,h_n)}{L_n} \equiv
\frac{\delta L_{n}}{L_{n}} - \left.\frac{\delta L_{n}}{L_{n}}\right|_{H=0} .
\end{align}
The magnetostriction depends on the magnetic field, $H$, only via
the effective field $h_n = g_n \mu_B H$. In the present
experimental setup the magnetic field has always been aligned
parallel to the measured length change such that the $g$-factor
also carries the same index $n$. The $g$-factors in the three
directions, the $a$ axis, $g_a$ (along the legs of the spin
ladder), the $b$ axis, $g_b$, and the $c^\star$ axis,
$g_{c^\star}$ (roughly along the ladder rungs), of \hp\ were
determined in Ref.~\onlinecite{Patyal90} to
\begin{align} \label{gfactors}
g_{a} = 2.06,\quad g_b = 2.18,\quad g_{c^\star} =  2.15.
\end{align}

The magnetostriction can be expressed in terms of the spin
correlators of the ladder,
\begin{align} \label{UniaxialMagnetoStr}
\frac{\Delta L_n(T,h_n)}{L_n} =
\gamma^{n}_\perp\,  \mathcal{D}_\perp(T,h_n) + \gamma^{n}_\parallel\, \mathcal{D}_\parallel(T,h_n),
\end{align}
where we introduced the abbreviation
\begin{align} \label{Dfunctions}
\mathcal{D}_{\perp,\parallel}(T,h) \equiv \mathcal{S}_{\perp,\parallel}(T,h) - \mathcal{S}_{\perp,\parallel}(T,0).
\end{align}
Similarly, the derivative of the magnetostriction,
\begin{align} \label{UniaxialMagnetoStr2}
\lambda_n(T,h_n) &=
\frac{1}{L_n}\frac{\partial \delta L_n}{\partial H}
= \gamma^{n}_\perp\,  \frac{\partial \mathcal{S}_\perp}{\partial H} + \gamma^{n}_\parallel\,
\frac{\partial \mathcal{S}_\parallel}{\partial H},
\end{align}
can be decomposed into derivatives of the correlators
$\mathcal{S}_{\perp,\parallel}$.

{\it Thermal expansion:} The uniaxial thermal expansion quantifies
the length changes of the crystal as a function of temperature at
constant magnetic field,
\begin{align}\label{UniaxialThermalExp}
\alpha_n(T,h_n) =
\left. \frac{1}{L_{n}} \frac{\partial \delta L_n}{\partial T}\right|_{H}
=
\gamma^{n}_\perp\,  \frac{\partial \mathcal{S}_\perp}{\partial T} + \gamma^{n}_\parallel\,
\frac{\partial \mathcal{S}_\parallel}{\partial T}.
\end{align}
It is determined by the derivatives of the correlators
$\mathcal{S}_{\perp,\parallel}$ with respect to temperature $T$.

\begin{figure}[b]
\includegraphics[width= 0.95\linewidth]{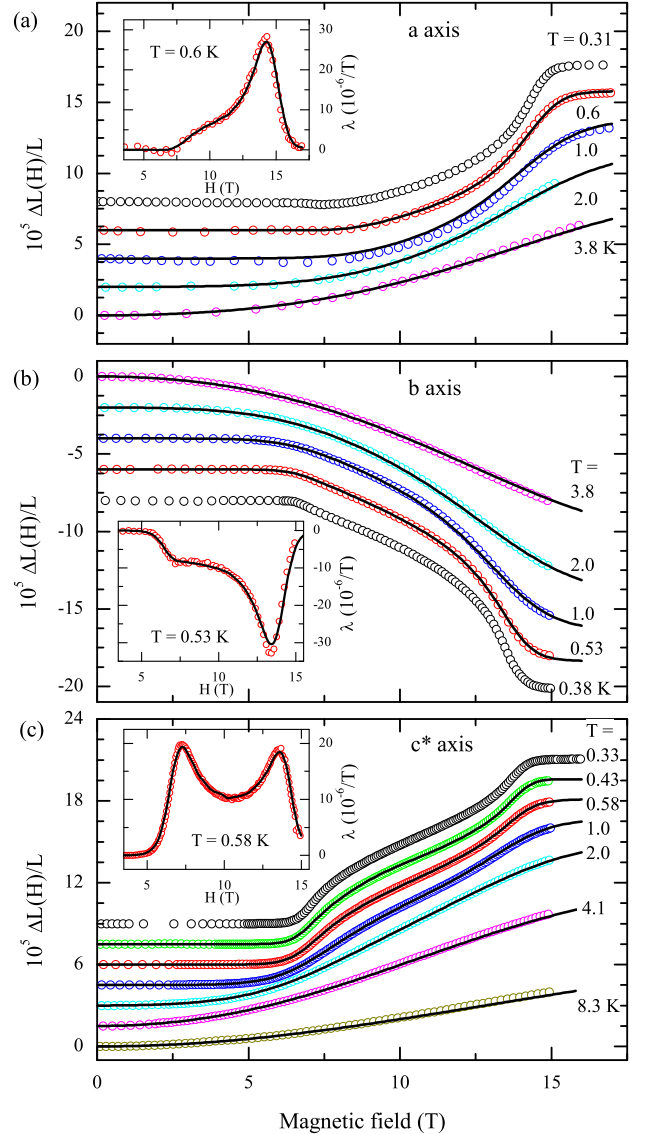}
\caption{\label{fig:comparison} (color online). The symbols in
each panel show the measured magnetostriction $\Delta L_n/L_n$ of
\hp\ along the $n$ axis with $n = a, b$, and $c^\star$ for
different temperatures. By construction, the magnetostriction
vanishes at $H=0$. The curves at different $T$ are off-set by $\pm
2\cdot 10^{-5}$ in panel (a) and (b), and by $1.5\cdot 10^{-5}$ in
panel (c). The solid lines are results from QMC simulations. The
insets show the fit to the $\lambda_n$ data from which the
$\gamma$ coefficients in Table~\ref{tab:GammaValues} have been
determined, see text in Sec.~\ref{theorycomparison}. }
\end{figure}

\subsection{Experimental results}
Uniaxial magnetostriction of \hp\ was measured in a range of
temperatures from $0.3$~K up to 8~K along the $a$, $b$ and
$c^{\star}$ axes, see panel (a), (b) and (c), respectively, of
Fig.~\ref{fig:comparison}. The measurements reveal a rich and
complex magnetoelastic behavior. From the low-temperature data,
the critical magnetic fields $H_{c1/2}$ for each direction can be
identified by the kinks in the magnetostriction. At small fields,
$H < H_{c1}$, and low temperatures, $T < J_\parallel/k_B$, the
magnetostriction is exponentially small as a consequence of the
presence of the spin-triplet gap. 
Near $H_{c1}$ the magnetostriction starts to grow and saturates
near the second critical field $H_{c2}$, becoming $H$ independent
for $H>H_{c2}$. For higher temperatures $T \gtrsim
J_{\parallel}/k_B$, the kinks near the critical fields are smeared
out. A saturation plateau at large fields is not reached anymore,
and the low-field behavior now becomes quadratic, $\Delta L_n/L_n
\sim H^2$. The inboxes show the derivatives $\lambda_n$, see
Eq.~(\ref{UniaxialMagnetoStr2}), for $T\approx 0.58\:{\rm K}\ll
J_{\parallel}/k_B$, that exhibit clear anomalies at both critical
fields $H_{c1}$ and $H_{c2}$. As discussed already in
Ref.~\onlinecite{ThomasPRL}, at lowest temperature the
magnetostriction and $\lambda_{c^\star}$ along the $c^\star$ axis
resemble closely the magnetization and susceptibility of \hp,
\cite{watson01a} respectively. For the measurements along the $a$
and $b$ directions, the shape of the magnetostriction derivative
is much less symmetric with respect to $(H_{c1}+H_{c2})/2$. In
particular, $\lambda_a$ has only a tiny peak at $H_{c1}$. As will
be seen below, this is due to the fact that the uniaxial pressure
dependence of $H_{c1}$ almost vanishes for $a$-axis pressure $p_a$
due to a partial cancellation of the uniaxial pressure
dependencies of $J_\perp $ and $J_\| $; see Eq.~(\ref{Hc1}) and
Table~\ref{tab:GammaValues}.

\begin{figure}
\includegraphics[width= 0.93\linewidth]{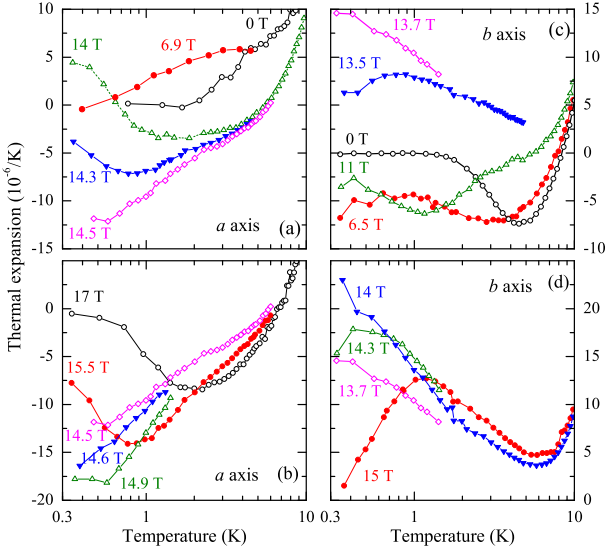}
\caption{\label{fig:alpha} (color online). Representative
measurements of the thermal expansion along the $a$ axis, i.e.\
the leg direction (left), and along the $b$ axis (right) of \hp .
The upper (lower) panels show data for magnetic fields below
(above) the upper critical field $H_{c2}$.}
\end{figure}

Figure~\ref{fig:alpha} displays some representative data of the
thermal expansion measurements along the $a$ axis (legs) and along
the $b$ axis. For both directions, we find a rich structure in
$\alpha_n$ with various sign changes, similar to our data along
the $c^\star$ axis which have been discussed in detail in
Ref.~\onlinecite{ThomasPRL}. Close to $H_{c2}$ there are again
clear indications for a diverging low-temperature thermal
expansion for both, $\alpha_a$ and $\alpha_b$. In contrast,
however, the features around $H_{c1}$ are much less pronounced,
which is due the aforementioned weak pressure dependencies of
$H_{c1}$ for $a$-axis and $b$-axis pressure.

\subsection{Two-parameter scaling}

\begin{figure}
\includegraphics[width= 0.95\linewidth]{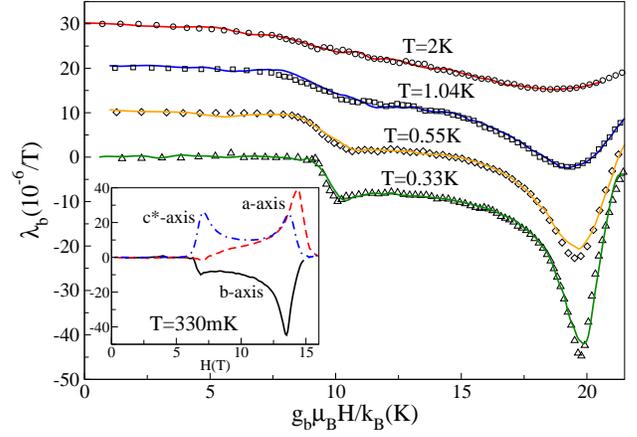}
\caption{\label{fig:2ParameterScaling} (color online). Uniaxial
magnetostriction along the $b$ axis, $\lambda_b$ see
Eq.~(\ref{UniaxialMagnetoStr2}), of \hp\ at different temperatures
(symbols). The higher temperature curves are each offset by
additional $10^{-5}/$T. The magnetic field is given in units of
Kelvin, $g_b \mu_B H/k_B$, with a $g$-factor along the $b$ axis
$g_b = 2.18$. The solid lines are fits
using the measurements along the $c^\star$ 
and $a$ axes 
and relation (\ref{LinearSuperposition}) with the
temperature-independent fitting parameters (\ref{FitParameters}).
The inset compares the magnetostriction along the three directions
at $T\simeq 330$~mK as a function of the magnetic field $H$. }
\end{figure}

A remarkable prediction of Eq.~(\ref{UniaxialMagnetoStr2}) is that
the magnetostriction curves along the three directions are
linearly dependent. In particular, the magnetostriction in one of
the three directions, say, along the $b$ axis can be obtained as a
linear superposition of the other two, along the $c^\star$ and $a$
axis, i.e.~the rung and leg directions, respectively,
\begin{align} \label{LinearSuperposition}
\lambda_b(T, h) = c_1 \lambda_c(T,h) + c_2 \lambda_a(T,h),
\end{align}
where $c_1$ and $c_2$ are constants independent of temperature and
magnetic field. It is important here that one accounts for the
$g$-factor anisotropy (\ref{gfactors}) by comparing the three
measurements at the same effective field $h = g \mu_B H$. The
sizeable $g$-factor anisotropy is reflected e.g.\ in the slight
shift of the positions of the high-field peaks of $\lambda_n$ for
the different field directions, as is illustrated in the inset of
Fig.~\ref{fig:2ParameterScaling}.

Fig.~\ref{fig:2ParameterScaling} confirms that the relation
(\ref{LinearSuperposition}) is obeyed by the experimental data.
The symbols in Fig. \ref{fig:2ParameterScaling} show the
measurements along the $b$ axis at different temperatures and the
solid line is obtained from the magnetostriction along the
$c^\star$ and $a$ axes and Eq.~(\ref{LinearSuperposition}) with
the fitted values
\begin{align} \label{FitParameters}
c_1 = -0.36, \quad c_2 = -0.85\: .
\end{align}
This agreement serves as an experimental proof that the
thermodynamics is fully grasped by a model Hamiltonian with only
two pressure-dependent coupling constants.

\subsection{Determination of the couplings $J_\perp$ and $J_\parallel$}
\label{determinationJ}
To perform a quantitative comparison of the dilatometric
experimental data with theory, we first need to determine the two
coupling constants, $J_\perp$ and $J_\parallel$, of the
spin-ladder Hamiltonian (\ref{modelhamiltonian}). For this
purpose, we take the critical fields as an experimental input and
derive $J_\perp$ and $J_\parallel$ from their second-order strong
coupling expressions\cite{Reigrotzki94}
\begin{align}
\label{Hc1} g \mu_B H_{c1} &= J_\perp \left( 1 -
\frac{J_\parallel}{J_\perp} + \frac{1}{2} \left(
\frac{J_\parallel}{J_\perp}\right)^2 +
\mathcal{O}\left(\frac{J_\parallel}{J_\perp}\right)^3 \right),
\\
\label{Hc2}
g \mu_B H_{c2} &= J_\perp + 2 J_\parallel.
\end{align}
The formula for $H_{c2}$ is exact in all orders in
$J_\parallel$.\cite{Chaboussant98}

\begin{figure}
\includegraphics[width= 1\linewidth]{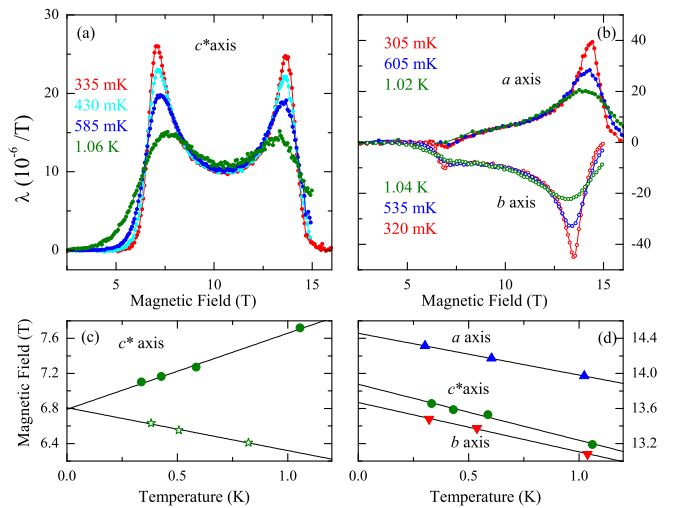}
\caption{\label{scaling1} (color online). The derivatives of the
magnetostriction, $\lambda_ {n}$ of \hp\ measured along
${c^\star}$ (panel a) and along the $a$ and $b$ axis (panel b) for
different temperatures. Panel (c) shows the linear-$T$ scaling of
the positions of the maxima of $\lambda_{c^\star}$ (dots) and of
the crossing points (stars) near $H_{c1}$ and (d) the positions of
the maxima of $\lambda_n$ near $H_{c2}$ for the different axes,
$n=a,b,c^\star$; see text for further explanation.}
\end{figure}

To obtain the values of $H_{c1}$ and $H_{c2}$ we make use of the
temperature scaling of the pronounced peaks near the critical
fields in the derivative of the magnetostriction along the
$c^\star$ axis, $\lambda_{c^\star}(H)$, see Fig.~\ref{scaling1}a.
As the temperature is lowered, the peak positions are expected to
approach the critical magnetic fields. From the known universality
class of the quantum phase transitions, cf.\ the discussion in
Sec.~\ref{sec:QuantumCriticality}, we expect the magnetic field
value of the peak positions to scale linearly with temperature
close to criticality, and this is indeed confirmed by the
experimental data as shown in Fig.~\ref{scaling1}c and d
(circles). From the extrapolation we obtain for the critical
fields along the $c^\star$ axis, $H^{c^\star}_{c1}=6.8$~T and
$H^{c^\star}_{c2}=13.9$~T. These values correspond to critical
effective fields, $h_{c1/2} = g_{c^\star} \mu_B
H^{c^\star}_{c1/2}$,
\begin{align} \label{CriticalFields}
h_{c1}/k_B=9.8\:{\rm K},\quad \quad h_{c2}/k_B=20.1\:{\rm K}\,.
\end{align}
The measurements along $a$ and $b$ yield practically (within
0.1~K) the same value for $h_{c2}$, but due to the small peaks
around $H_{c1}$ they do not allow for a reliable quantitative
determination of the lower critical fields.

Another feature of the peaks in $\lambda_{c^\star}(H)$ is the
steepening of its left and right flank near $H_{c1}$ and $H_{c2}$,
respectively, upon lowering $T$. As a consequence, two $\lambda_
{c^\star}(H)$ measurements closest in temperature show a crossing
point on these flanks whose position approaches the critical
magnetic fields as the mean of their temperatures vanishes. The
extrapolation of the crossing point position confirms the
extracted value of $H^{c^\star}_{c1}$ as shown in
Fig.~\ref{scaling1}c (stars). The corresponding test for the
$H^{c^\star}_{c2}$ value is not possible due to the enhanced noise
in the experimental $\lambda_ {c^\star} $ data at higher magnetic
fields.

From the critical fields of the $c^{\star}$-axis measurements we
obtain the coupling constants by using Eqs.~(\ref{Hc1}) and
(\ref{Hc2}) that are valid up to second order in the small
parameter $J_\parallel/J_\perp$. We get
\begin{align}\label{physicalpar}
J_\perp/k_B = 12.9\:{\rm K},\quad J_\parallel/ k_B = 3.6\:{\rm
K}\,.
\end{align}
These values are used for the Quantum Monte Carlo simulations of
the spin-ladder Hamiltonian (\ref{modelhamiltonian}) presented in
the following.

\subsection{Quantitative comparison of magnetostriction with theory}
\label{theorycomparison}
\begin{figure}
\includegraphics[width= \linewidth]{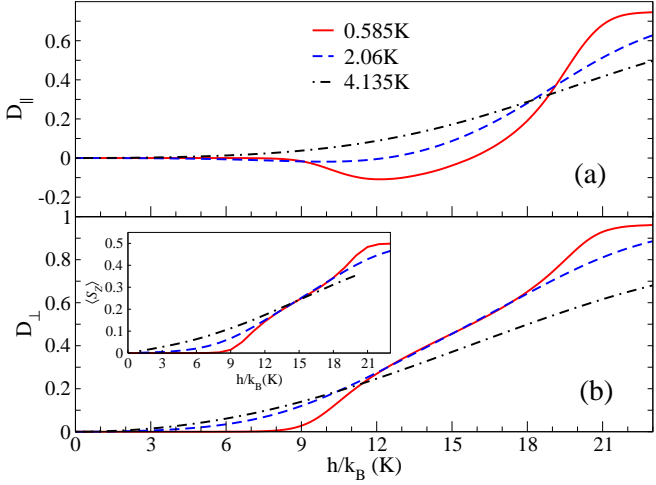}
\caption{\label{fig:correlators} (color online). Correlators
$\mathcal{D}_{\perp,\parallel}$, see Eq.~(\ref{Dfunctions}), of
the spin ladder Hamiltonian (\ref{modelhamiltonian}) with the
parameters (\ref{physicalpar}), computed with QMC as a function of
the effective magnetic field $h = g \mu_B H$ (in units of Kelvin)
for different temperatures. The simulations are performed always
in the limit of large number of sites $N \gg J_\perp/(k_B T)$. The
inset shows the magnetization per spin of the ladder. Note the
similarity between the rung correlator and the magnetization at
low $T$. }
\end{figure}

With the values for the exchange couplings (\ref{physicalpar}) the
spin-ladder Hamiltonian describing \hp\ is determined. We computed
with the Quantum Monte Carlo (QMC) algorithm\cite{Olav00} the spin
correlation functions that enter the dilatometric quantities. In
Fig.~\ref{fig:correlators} we show the numerically evaluated
correlators $\mathcal{D}_{\parallel}$ and $\mathcal{D}_{\perp}$ of
Eq.~(\ref{Dfunctions}) as a function of the effective magnetic
field, $h = g \mu_B H$, for a series of different temperatures.

The rung correlator $\mathcal{D}_\perp$ in
Fig.~\ref{fig:correlators}b increases monotonically with
increasing $h$. This can be simply understood in the
strong-coupling limit, $J_\perp \gg J_\parallel$. The rung dimers
then form an almost perfect singlet at $h=0$ such that a finite
magnetic field augments the triplet component to the ground state
and thus enhances the ferromagnetic correlations. At low
temperatures, the rung correlator $\mathcal{D}_\perp$ effectively
measures the density of triplets and attains a shape very similar
to the magnetization curve of the ladder, shown in the inset of
Fig.~\ref{fig:correlators}b. In zero field, $h=0$, the correlator
$\mathcal{D}_\perp$ vanishes by construction while at $h \to
\infty$ it saturates to a value close to one. The deviation from
one is due to the in-chain quantum fluctuations at $h=0$ that
reduce the antiferromagnetic correlations along the rung. We can
obtain the saturation value from the following expansion for the
ground state energy per rung \cite{Reigrotzki94}
\begin{align}\label{GSenergy}
E_0 = - J_\perp \left(
\frac{3}{4} + \frac{3}{8}\left(\frac{J_\parallel}{J_\perp}\right)^2
+ \frac{3}{16}\left(\frac{J_\parallel}{J_\perp}\right)^3 +
\mathcal{O}\left(\frac{J_\parallel}{J_\perp}\right)^4
\right).
\end{align}
For the parameters (\ref{physicalpar}) we get for the saturation value
\begin{equation}   \label{LimitPerpCorrelator}
\left.\mathcal{D}_\perp(0,h)\right|_{h\to \infty} =
\frac{1}{4} - \frac{\partial E_0}{\partial J_\perp} \approx 0.96
\end{equation}
in agreement with our numerical findings in Fig.~\ref{fig:correlators}b.

At low temperatures, the leg correlator $\mathcal{D}_\parallel$ of
Fig.~\ref{fig:correlators}a is instead a non-monotonic function of
the magnetic field. The existence of a minimum can be understood
in terms of competing Resonant Valence Bonds.\cite{anderson87}
Below the first critical field, the ground state will consist of
singlet bonds that for $J_\perp \gg J_\parallel$ are mainly formed
along the rungs. For $h \gtrsim h_{c1}$, the ground state acquires
a small triplet component, that is reflected in the presence of
rare rung-triplet states. In the presence of these triplets,
stronger singlets can be accommodated along the chains enhancing
effectively the antiferromagnetic correlations along the legs such
that $\mathcal{D}_\parallel$ first decreases. However, increasing
$h$ even further the density of triplets grows, their singlet
screening clouds start to overlap, and ferromagnetism finally
prevails leading to a sign change in the $\mathcal{D}_\parallel$
function.

The saturation value at $T=0$ can again be derived with the help
of the ground state energy (\ref{GSenergy}),
\begin{align} \label{LimitParallelCorrelator}
\left.\mathcal{D}_\parallel(0,h)\right|_{h\to \infty} &=
\frac{1}{2} - \frac{\partial E_0}{\partial J_\parallel} \approx 0.75,
\end{align}
that coincides with the numerical value in Fig~\ref{fig:correlators}a.

In Figs.~\ref{fig:comparison}, we show the comparison between the
experimentally measured magnetostrictions $\Delta L_n/L_n$ and the
Quantum Monte Carlo data. With the help of the correlators
$\mathcal{D}_\perp$ and $\mathcal{D}_\parallel$ computed for $T=
580$~mK as a function of $H$ and the relation
(\ref{UniaxialMagnetoStr}), we determine the mixing coefficients,
$\gamma^n_\perp$ and $\gamma^n_\parallel$ of
Eq.~(\ref{MixingCoefficients}), by a fit to the experimental data.
The resulting values of $\gamma^n_{\perp,\parallel}$ are listed in
Table~\ref{tab:GammaValues}.
\begin{table}[t]
\begin{tabular}{cccrr}
axis  & $\gamma^n_\perp \times 10^{5}$ & $\gamma^n_\parallel
\times 10^{5}$ &
$\partial \ln J_\perp/\partial p_n$ & $\partial \ln J_\parallel/\partial p_n$ \\
\tableline
$a$ & 4.2 & 7.7 & 20\%/GPa & 133\%/GPa\\
$b$ & -8.2 & -6.0 & -40\%/GPa& -104\%/GPa\\
$c^\star$ & 13.5 & -1.1 & 65\%/GPa& -19\%/GPa
\end{tabular}
\caption{\label{tab:GammaValues} Magnetoelastic couplings
$\gamma^n_\alpha = (\partial J_\alpha/\partial p_n)/V_D$, see
Eq.~(\ref{MixingCoefficients}), obtained from a fit of the
magnetostriction of \hp\ to the QMC data as shown in
Fig.~\ref{fig:comparison}, and the resulting uniaxial pressure,
$p_n$, dependencies of the exchange couplings $J_\perp$ and
$J_\parallel$. The changes of $J_\perp$ and $J_\parallel$ under
hydrostatic pressure are given by the respective sums of their
uniaxial pressure dependencies.}
\end{table}
One can check that the coefficients in Table~\ref{tab:GammaValues}
are consistent with the relation (\ref{LinearSuperposition}) using
(\ref{FitParameters}). Remarkably, these coefficients determined
at one temperature yield parameter-free predictions for all the
magnetostriction curves measured at other temperatures. As shown
in Fig.~\ref{fig:comparison}, the agreement between the theory and
experiment is perfect in a range of $T$ of almost one decade. The
experimental curves at $T<400$~mK have not been calculated because
our QMC algorithm could not produce reliable data at such low
temperature.

The two correlators $\mathcal{D}_{\perp,\parallel}$ contribute
roughly equally to the magnetostriction along the $a$ axis (leg
direction) and the $b$ axis; the corresponding ratio of mixing
coefficients is of order one. Along the ${c^\star}$ axis (rung
direction), however, the magnetostriction is dominated by
$\mathcal{D}_\perp$ except for a small $8\%$ admixture of the leg
correlator. This agrees with the naive expectation that squeezing
the ladder along the rung mainly affects $J_\perp$. As a
consequence, the magnetostriction $\Delta L_{c^\star}/L_{c^\star}$
inherits the characteristic shape of the correlator
$\mathcal{D}_\perp$ that resembles at low $T$ the magnetization of
the ladder as a function of $H$. This also explains the similarity
between the susceptibility\cite{watson01a} of \hp\ and the
derivative $\lambda_{c^\star}$, that shows two characteristic
peaks located close to the critical fields $H_{c1/2}$. In
Ref.~\onlinecite{ThomasPRL}, we neglected the $8\%$ admixture of
the leg correlator and interpreted the thermal expansion data
along the $c^\star$ direction solely in terms of a pressure
dependence of $J_\perp$. In this approximation, the thermal
expansion $\alpha_{c^\star}$ can be related to the $H$ derivative
of the entropy, $\alpha_{c^\star} \propto \partial S/\partial H$,
that allows to understand in simple terms the various sign changes
of thermal expansion.\cite{Garst05}

\section{Quantum criticality}
\label{sec:QuantumCriticality}

Near the two critical fields, $H_{c1}$ and $H_{c2}$, the
lowest-lying triplet excitation or the spin-wave excitation become
gapless, respectively. Both transitions are described by a
Hamiltonian of free non-relativistic fermions, $c(x)$,
representing the corresponding
excitations\cite{Chitra96,Sachdev94}
\begin{align} \label{CriticalHamiltonian}
\mathcal{H}_{\rm cr} = \int dx\, c^\dag(x)
\left(- \frac{\hbar^2\partial^2_x}{2 m} + r \right) c(x)
\end{align}
where the mass $m$ differs for the two transitions and the control
parameter, $r \propto H-H_{c1/2}$, measures at $T=0$ the distance
to the quantum critical point. The resulting critical contribution
to the free energy density is
\begin{align} \label{CrFreeEnergy}
f_{\rm cr}(r,T) &= - \frac{a}{V_D} \frac{\sqrt{2 m}}{\hbar} (k_B
T)^{3/2} \mathcal{F}\left(\frac{r}{k_B T}\right)
\end{align}
with the volume $V_D$ per rung and the distance $a$ between the
rungs. The scaling function is given by
\begin{align} \label{Ffunction}
\mathcal{F}(x) &= \int_0^\infty \frac{d y}{\pi}
\log\left[ 1 + \exp\left(-y^2 - x\right)\right].
\end{align}

\subsection{Critical thermal expansion near $H_{c2}$}
\label{sec:CriticalThermalExp}

In the following, we will focus on the critical behavior near the
upper critical field, $H_{c2}$. The ground state of the system for
$H>H_{c2}$ is particularly simple: all spins are fully polarized
by the magnetic field. This allows to determine the parameters
exactly,\cite{Chaboussant98}
\begin{align}
\label{MassHc2} m &= \frac{\hbar^2}{J_\parallel a^2}, \quad r = g
\mu_B H - (J_\perp + 2 J_\parallel)\: .
\end{align}
We will consider the predictions for the uniaxial thermal
expansion $\alpha_n$ and compare it to experiment. The uniaxial
length change close to $H_{c2}$ can be obtained from
(\ref{MeanStrain}) by approximating the magnetic free energy with
its critical part, $F_m/V \approx f_{\rm cr}$. Moreover, the
dominant contribution originates from the sensitivity of $f_{\rm
cr}$ upon small variations of the control parameter $r$ so that we
can neglect $\partial f_{\rm cr}/\partial m$  as it only gives
sub-leading corrections close to the quantum critical point. So we
obtain
\begin{align}
\frac{\delta L^{\rm cr}_{n}}{L_{n}} = - V_D
\frac{\partial f_{\rm cr}}{\partial r} \left(\gamma^n_{\perp}
+ 2 \gamma^n_{\parallel}\right) .
\end{align}
From this expression, we derive the critical uniaxial thermal
expansion close to $H_{c2}$, $\alpha^{\rm cr}_n = (\partial \delta
L^{\rm cr}_{n}/\partial T)/L_n$,
\begin{align} \label{CriticalAlpha}
\alpha_n^{\rm cr} = \left(\gamma^n_{\perp} + 2 \gamma^n_{\parallel} \right)
\sqrt{\frac{2 k_B}{J_{\parallel} T}}
\left[\frac{1}{2} \mathcal{F}'(x) -
x \mathcal{F}''(x)\right]_{\textstyle x=\frac{r}{k_B T}}
\end{align}
where the primes indicate derivatives of the function
$\mathcal{F}(x)$, (\ref{Ffunction}). Formula (\ref{CriticalAlpha})
is the limiting behavior of thermal expansion near the quantum
critical point at $H_{c2}$. It is expected to describe
asymptotically the experimental data in the range $|r|/k_B,\, T
\ll J_\parallel/k_B = 3.6$~K. At criticality, $r=0$, the thermal
expansion diverges with temperature as
$1/\sqrt{T}$,\cite{ThomasPRL}
\begin{align} \label{CriticalAlpha2}
\left.\alpha_n^{\rm cr}\right|_{H=H_{c2}} = \mathcal{C}
\left(\gamma^n_{\perp} + 2 \gamma^n_{\parallel} \right)
\sqrt{\frac{k_B}{J_{\parallel} T}},
\end{align}
with the prefactor $\mathcal{C} = \left(\sqrt{2} - 1 \right)
\zeta(1/2)/(2 \sqrt{2 \pi}) \approx -0.12066$. The next-to-leading
order correction to the result (\ref{CriticalAlpha2}) originates
from two-magnon scattering processes that lead to a
temperature-independent contribution to $\alpha_n$ at $H=H_{c2}$.
The strain dependence of the mass $m$ in (\ref{MassHc2}) and
higher-order derivatives of the magnon dispersion, on the other
hand, yield corrections to (\ref{CriticalAlpha2}) that are
suppressed by a relative factor $k_B T/J_\parallel$. All these
corrections to scaling are neglected in the following.

\begin{figure}
\includegraphics[width= \linewidth]{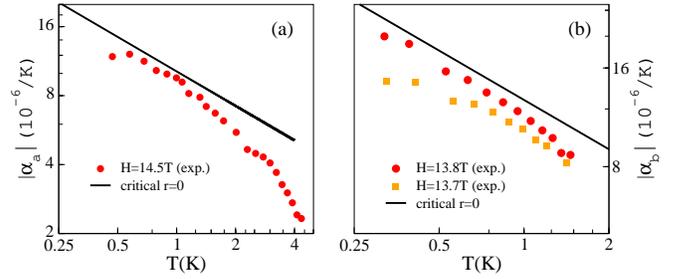}
\caption{\label{fig:criticalAlpha} (color online). Quantum
critical thermal expansion near $H_{c2}$ of \hp\ (symbols) along
the $a$ and $b$ axis and comparison with the parameter-free theory
(\ref{CriticalAlpha2}) (lines). The critical fields along the
three directions as extracted in Fig.~\ref{scaling1}d are given by
$H^a_{c2} = 14.5$~T and $H^b_{c2} = 13.7$~T.}
\end{figure}

The mixing coefficients $\gamma_{\perp,\parallel}^n$ entering
(\ref{CriticalAlpha}) have been already determined in
Sec.~\ref{sec:Dilatometry} and are listed in
Table~\ref{tab:GammaValues}. This allows us to perform a
parameter-free calculation of $\alpha_n$ near the quantum critical
point at $H_{c2}$. Fig.~\ref{fig:criticalAlpha} shows a comparison
of the uniaxial thermal expansion of \hp\ with formula
(\ref{CriticalAlpha2}) (lines) on a double-logarithmic scale. Note
that the critical field $H_{c2}$ differs for the three
crystallographic axes, as it is given by $H^n_{c2} = h_{c2}/(g_{n}
\mu_B)$ where $h_{c2}/k_B = 20.1$~K and the $g$-factors are listed
in Eqs.~(\ref{gfactors}). Measurements of $\alpha_a$ were
performed at the critical field $H^a_{c2} = 14.5$T and are
displayed in Fig.~\ref{fig:criticalAlpha}a. The data nicely
approaches the asymptotic $1/\sqrt{T}$ behaviour at lowest
temperatures. Panel (b) displays $\alpha_b(T)$ along the $b$ axis
for two fields closest to the critical field. The saturation of
the curve at $H=13.7$~T at lowest temperatures suggests that the
critical field is  slightly higher than the value
$H^b_{c2}=13.7$~T which follows from Eq.~(\ref{CriticalFields})
with $g$-factor $g_b = 2.18$. The deviation is, however, within
the error bar for the critical field value of $\simeq \pm 0.05$~T
associated with the analysis in Fig.~\ref{scaling1}d. As expected,
in both panels the corrections to scaling behaviour of $\alpha_n$
are still sizeable in the considered temperature range.

\subsection{First-order transition driven by quantum criticality}
\label{sec:FirstOrder}

The above considerations were based on the assumption made in
Sec.~\ref{sec:Dilatometry} that the elastic tensor
(\ref{ElasticTensor}) is independent of magnetic field and
temperature and, in particular, is such to ensure the stability of
the crystal. Here we show that this assumption breaks down in the
immediate vicinity of the quantum critical points of the spin
ladder. Close to the quantum phase transition, the spin ladder is
described by the Hamiltonian (\ref{CriticalHamiltonian}). Near
criticality, the most relevant magnetoelastic interaction derives
from the strain dependence of the control parameter $r$. As $r$ is
just a certain function of the exchange couplings $J_\alpha$, we
can repeat the reasoning of Sec.~\ref{sec:Dilatometry} to derive
an effective magnetoelastic interaction Hamiltonian,
\begin{align} \label{CrET}
\mathcal{H}^{\rm cr}_{\rm int} = \int dx\, c^\dag(x) c(x) g^{ij}_r u_{ij}(x)
\end{align}
where the coupling $g^{ij}_r = \partial r/\partial u_{ij}|_{u=0}$
measures the strain dependence of the control parameter. In
second-order perturbation theory in $g$, this interaction then
leads to an effective elastic tensor of the form,
\begin{align} \label{CriticalET}
c_{ijkl} = c^0_{ijkl} - \chi_{\rm cr}\, g^{ij}_r g^{kl}_r .
\end{align}
The strong temperature and magnetic field dependence enters via
the susceptibility $\chi_{\rm cr}$ that is obtained from the
critical free energy (\ref{CrFreeEnergy})
\begin{align} \label{CriticalSusc}
\chi_{\rm cr} =  - \frac{\partial^2 f_{\rm cr}}{\partial^2 r} =
\frac{a}{V_D} \frac{\sqrt{2 m}}{\hbar \sqrt{k_B T}}
\mathcal{F}''\left(\frac{r}{k_B T}\right) .
\end{align}
Note that for the spin ladder, $r \propto H - H_{c1/2}$, such that
$\chi_{\rm cr}$ indeed coincides with the critical magnetic
susceptibility of the system. Moreover, $\chi_{\rm cr}$ is
positive so that the magnetoelastic coupling correction in
(\ref{CriticalET}) reduces the elastic moduli. In fact, the
susceptibility $\chi_{\rm cr}$ diverges -- like the thermal
expansion -- as $1/\sqrt{T}$ at criticality, $r=0$, resulting in a
strong softening of the crystal until it becomes unstable
sufficiently close but still away from the quantum critical point.
The lattice is then expected to undergo a first-order transition
that preempts quantum criticality.

The singular contribution to the elastic moduli in (\ref{CrET}) is
attributed to the low dimensionality $d=1$ of the critical system,
i.e., the spin ladder. Generally, in the presence of a strain
coupling to the square of the order parameter, a divergent
contribution to elasticity is expected when
\begin{align} \label{1stOCondition}
\nu (d+z) < 2,
\end{align}
where $\nu$ is the correlation length and $z$ is the dynamical
exponent of the quantum critical point. If this criterion is
fulfilled the lattice becomes unstable before the quantum phase
transition is reached. The condition (\ref{1stOCondition}) is
related to the corresponding criterion for compressible classical
critical systems.\cite{LandauBook,Rice,Domb,LarkinPikin} There, it
is known that a second-order transition is preempted by an
instability of the lattice if the specific heat exponent is
positive, $\alpha = 2 - \nu d > 0$. From this, the criterion for
compressible quantum critical systems (\ref{1stOCondition}) is
obtained by replacing the spatial dimension $d$ with the effective
dimension $d+z$.

Compressible critical systems have been studied extensively in the
past, see Ref.~\onlinecite{BergmanHalperin76} and references
therein. Renormalization group (RG) treatments of such systems
yield run-away RG flow that is interpreted as an indicator for a
first-order transition. For isotropic media, where the elastic
tensor is characterized by just two moduli, compression and shear,
this conclusion is borne out by explicit
calculations.\cite{LarkinPikin} Interestingly, the arising
transition turns out to be governed by long-range interactions and
is therefore mean-field like. An analysis of compressible critical
models for materials with lower crystal symmetry is in general,
however, rather
challenging.\cite{BergmanHalperin76,deMoura76,Cowley76}

Here, we do not aim for a complete description of a possible
first-order transition in the monoclinic crystal structure of \hp\
triggered by quantum critical fluctuations. For an estimate of the
location of first-order transitions in the phase diagram, we
consider instead the elastic properties on the level of a
simplistic mean-field theory. Neglecting phonon excitations and
crystal anisotropies, an effective Landau  potential for the
macroscopic volume change $u = \Delta V/V$ can be derived
\begin{align} \label{LandauPot}
\mathcal{V}(u) = \frac{K_{\rm eff}}{2} u^2 + f_{\rm cr}( r +g u, T)
\end{align}
where $K_{\rm eff}$ is an effective bulk modulus and with $f_{\rm
cr}$ given in Eq.~(\ref{CrFreeEnergy}). The first argument of
$f_{\rm cr}$ is the control parameter expanded in first order in
$u$ with the coupling constant $g \equiv \left.\partial r/\partial
u\right|_{u=0}$. For conditions of fixed hydrostatic pressure,
minimization of the potential $\mathcal{V}$ yields the value of
volume change $u_{\rm min}(r_0,T)$ at given $r_0$ and temperature
$T$.  At $r_0 = T = 0$, the critical part behaves as $f_{\rm cr}
\sim |u|^{\nu (d+z)} \Theta(-u)$ and dominates over the quadratic
part of the potential if the criterion  (\ref{1stOCondition}) is
fulfilled; for the spin ladder we have $\nu(d+z) = 3/2$. Due to
this strong non-analyticity of the Landau potential
(\ref{LandauPot}) near the apparent quantum critical point, the
value of $u_{\rm min}$ jumps in its vicinity and thus avoids the
strong critical fluctuations by prohibiting the argument $r_0 + g
u_{\rm min}$ to approach zero.

\begin{figure}
\includegraphics[width= 0.8\linewidth]{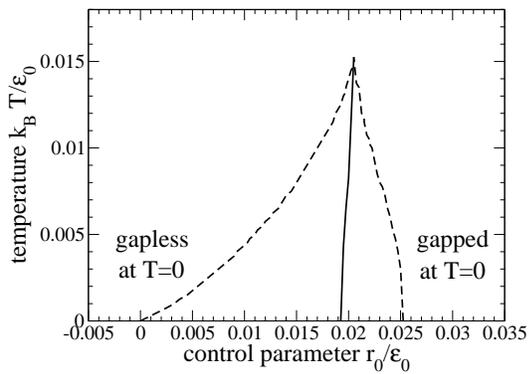}
\caption{\label{fig:FirstOrder} Phase diagram numerically
evaluated from the Landau potential (\ref{LandauPot}); temperature
and control parameter are given in units of the energy scale
$\varepsilon_0 = \frac{g^4 a^2 2 m}{V_D^2 K_{\rm eff}^2 \hbar^2}$.
The first-order transition line (solid) is embedded in a
coexistence region which is bounded by the dashed lines; the
critical fluctuations at $T=0$ are gapped (gapless) on the right
(left) hand side of the transition line. }
\end{figure}

The resulting line of first-order transitions and the coexistence
region in the phase diagram are shown in
Fig.~\ref{fig:FirstOrder}. The line of first-order transitions
terminates in a second-order endpoint, which we numerically
evaluated to be located at
\begin{align} \label{endpoint}
r^*/\varepsilon_0 = 0.0205 \pm 0.0003,\quad
k_B T^*/\varepsilon_0 = 0.0155 \pm 0.0005.
\end{align}
The characteristic energy scale $\varepsilon_0 = \frac{g^4 a^2 2
m}{V_D^2 K_{\rm eff}^2  \hbar^2}$ is suppressed by the fourth
power of the coupling $g$ and the temperature $T^*$ is therefore
expected to be small. We expect the second-order transition at the
endpoint $(r^*,T^*)$ to be particular: generically, it should be
mean-field like and as such it is not accompanied by critical
fluctuations.\cite{LarkinPikin,Cowley76}

In order to get the order of magnitude for the transition
temperature $T^*$ we estimate the energy scale $\varepsilon_0$
near the critical field $H_{c2}$ of \hp. Away from criticality,
the coupling $g$ obeys $g = \frac{\partial}{\partial u} r = -
K_{\rm eff} \frac{\partial}{\partial p} r \equiv K_{\rm eff} V_D
\gamma$; with the expression for the mass $m$,
Eq.~(\ref{MassHc2}), we then obtain $\varepsilon_0 =  \frac{2
\gamma^4 V_D^2 K_{\rm eff}^2}{J_\parallel}$. Using for the modulus
the estimate $K_{\rm eff} \sim 10$ GPa (assuming a rather soft
material) and a $\gamma$ value, $\gamma \sim 10^{-4}$, of the
order of the ones in Table~\ref{tab:GammaValues} we get the rough
estimate $\varepsilon_0/k_B \sim 10^{-5}$~K. The small numerical
prefactor in (\ref{endpoint}) suppresses the critical temperature
further by two orders of magnitude, $T^* \sim 10^{-7}$~K. This
small value of $T^*$ suggests that the phenomenon of
fluctuation-induced first-order transitions close to quantum
criticality is irrelevant for \hp\ and cannot be observed. In
particular, the above discussion applies only in the temperature
regime where thermodynamics is dominated by the one-dimensional
physics of the spin ladders. At criticality, the inter-ladder
coupling $J_{\rm 3d}$ leads to a dimensional crossover at a
temperature $T_{3d}$ from 1d to 3d quantum critical
behavior,\cite{Giamarchi99,Orignac07} which weakens the
non-analyticity in the effective Landau potential for the strain
$u$, and Eq.~(\ref{LandauPot}) ceases to be applicable for
temperatures $T \lesssim T_{3d}$. Moreover, quantum criticality in
the 3d regime is not strong enough to have the criterion
(\ref{1stOCondition}) fulfilled. In the framework of the 3d dilute
interacting Bose gas with anisotropic hopping,\cite{Giamarchi99}
the crossover temperature, $T_{3d}$, can be estimated to be of the
order of the inter-ladder coupling, $T_{3d} \sim J_{\rm 3d}/k_B$.
From the measured N\'eel temperature near $(H_{c1}+H_{c2})/2$,
$T_N \simeq 80$~mK,\cite{ruegg} it follows that $J_{\rm 3d}$ is at
least three orders of magnitude larger than our estimate of $T^*$.
This means that in \hp\ the divergent correction to the elastic
tensor (\ref{CrET}) is cut-off by inter-ladder interactions before
any modulus vanishes. Nevertheless, the strong softening of the
crystal close to the critical fields should give pronounced
signatures, e.g., in ultrasound measurements.

The derivation of the quantum critical correction to the elastic
tensor (\ref{CriticalET}) was based on very general arguments. In
particular, the prediction of the correction being proportional to
the magnetic susceptibility $\chi$ carries over to any
magnetic-field driven quantum critical point. Such a
correspondence between $\chi$ and the elastic moduli has been
indeed found by Schmidt {\it et al.}\cite{Schmidt01} in the
spin-dimer system NH$_4$CuCl$_3$. This compound exhibits a series
of quantum phase transitions as a function of magnetic field that
are ascribed to the successive polarization of nearly decoupled
spin-dimer subsystems.\cite{Ruegg04,Matsumoto03}

\section{Summary}

The magnetostriction and thermal expansion of \hp\ are described
to a remarkable level of accuracy by the two-leg spin-ladder
Hamiltonian of Eq.~(\ref{modelhamiltonian}), with the two exchange
couplings $J_\perp$ and $J_\parallel$ given in
Eq.~(\ref{physicalpar}). The presence of only two characteristic
energy scales leads to an interdependence of the magnetostriction
measured along different crystallographic directions. This
predicted dependence is confirmed by the experiments, as shown in
Fig.~\ref{fig:2ParameterScaling}, and justifies\cite{watson01a}
the disregard of
other possible terms in the Hamiltonian, e.g. diagonal or ring
exchange couplings. As a consequence, the contribution of the spin
subsystem to the elastic properties are captured by two static
nearest-neighbor spin-spin correlation functions,\cite{Zapf07}
which we calculated numerically with QMC. Their contribution to
magnetostriction along some arbitrary axis $n$ is weighted by
effective magnetoelastic coupling constants that we denoted by
$\gamma^n_{\perp,\parallel}$, see Eq.~(\ref{MixingCoefficients}).
A detailed fit to the experimental magnetostriction data, see
Fig.~\ref{fig:comparison}, allowed us to determine these coupling
coefficients for the three crystallographic directions $a$, $b$
and $c^\star$ of \hp. Their values are given in
Table~\ref{tab:GammaValues}.

In particular, we find that along the rungs of the ladder, i.e.
the $c^\star$ axis, the magnetostriction is dominated by the
strain dependence of the rung coupling $J_\perp$. This explains
the resemblance of the magnetostriction along the $c^\star$ axis
with the magnetization of the ladder.\cite{watson01a} This
correspondence was already exploited in
Ref.~\onlinecite{ThomasPRL} to explain the three consecutive sign
changes of the thermal expansion along $c^\star$, that are
observed at lowest temperatures as a function of magnetic field,
in terms of entropy extrema.\cite{Garst05}

In addition, we analyzed the thermal expansion near the upper
critical field $H_{c2}$. Sufficiently close to the quantum
critical point at $H_{c2}$, the spin-ladder Hamiltonian
(\ref{modelhamiltonian}) becomes equivalent to a model of free
non-relativistic fermions with known microscopic parameters, see
Sec.~\ref{sec:CriticalThermalExp}. This enables us to derive a
parameter-free analytic formula for the thermal expansion along
the three crystallographic directions. At criticality, $H=H_{c2}$,
we observe a $1/\sqrt{T}$ divergence of thermal expansion down to
our lowest temperatures. As we discussed in detail in
Ref.~\onlinecite{ThomasPRL}, this strong singularity is rooted in
the low dimensionality of the critical subsystem.

Finally, we predict a correction to the elastic moduli of \hp,
whose magnetic field and temperature dependence is governed by
static four-spin correlation functions of the spin ladder.
Typically, this correction is small, and it is indeed negligible
for our analysis of magnetostriction in the measured temperature
range. However, close to quantum criticality, the magnetic
correction to the elastic moduli becomes proportional to the
magnetic susceptibility $\chi$. As the susceptibility diverges at
criticality as $\chi \sim 1/\sqrt{T}$, this should lead to a
strong softening of the crystal rendering the elastic system in
principle unstable at sufficiently low temperatures, see
Sec.~\ref{sec:FirstOrder}. We argued that this phenomenon is in
fact generic for quantum critical systems if the strain couples to
the square of the order parameter and, in addition, the criterion
$\nu (d+z) < 2$ is fulfilled.  Our estimates show, however, that
in \hp\ this fluctuation induced first-order transition is not
realized as the small inter-ladder coupling will cut-off the
strong singularities before an instability can develop.

Whereas the strong singularity of $\chi$ is particular to the
quantum critical point of the spin ladder, a correction to the
elastic tensor proportional to the magnetic susceptibility is a
generic feature of magnetic-field driven quantum criticality. Such
corrections have been observed in ultrasound experiments on
NH$_4$CuCl$_3$ by Schmidt {\it et al.}.\cite{Schmidt01}

As an outlook, we remark that the precise knowledge of the
magnetoelastic couplings of \hp\ as given in
Table~\ref{tab:GammaValues} will be especially useful in the quest
for a theoretical explanation of its thermal transport
properties.\cite{ThermalTransport}

\acknowledgements

We acknowledge useful discussions with T.~Giamarchi, J.A.~Mydosh,
K.~Kiefer, A.~Sologubenko, B.~Thielemann, and M.~Vojta. This work
was supported by the Deutsche Forschungsgemeinschaft through
Sonderforschungsbereich~608 and by the Swiss National Science
Foundation. Computer time was allocated through the Swedish Grant
No.~SNIC 005/06-8.


\end{document}